\documentclass[12pt]{article}

\pdfoutput=1

\usepackage{latexsym,amsmath,amssymb,amsfonts,amsthm,bbm,mathrsfs,breakcites,dsfont,xcolor}
\usepackage{stmaryrd}
\usepackage[ruled, vlined]{algorithm2e}
\usepackage{epsfig}
\usepackage{natbib}
\usepackage{url}
\usepackage{dsfont}
\usepackage{enumerate}
\usepackage{graphicx}
\usepackage{graphics}
\usepackage{caption}

\usepackage{color}
\usepackage{dcolumn}
\usepackage{bm}
\usepackage{float}
\usepackage{hyperref} 
\hypersetup{ colorlinks = true, citecolor = black, urlcolor = blue}

\definecolor{background-color}{gray}{0.98}

\newtheorem{theorem}{Theorem}

\DeclareMathOperator*{\sargmin}{sargmin}

\title{Random projections: data perturbation for classification problems}

\author{Timothy I. Cannings \\ \textit{School of Mathematics, University of Edinburgh}}

\date{}
\begin{document}
\maketitle

\begin{abstract}
\noindent Random projections offer an appealing and flexible approach to a wide range of large-scale statistical problems. They are particularly useful in high-dimensional settings, where we have many covariates recorded for each observation.  In classification problems there are two general techniques using random projections.  The first involves many projections in an ensemble -- the idea here is to aggregate the results after applying different random projections, with the aim of achieving superior statistical accuracy.  The second class of methods include hashing and sketching techniques, which are straightforward ways to reduce the complexity of a problem, perhaps therefore with a huge computational saving, while approximately preserving the statistical efficiency.  
\end{abstract}
\begin{figure}[h!]
	\centering
	\includegraphics[width=0.75\textwidth]{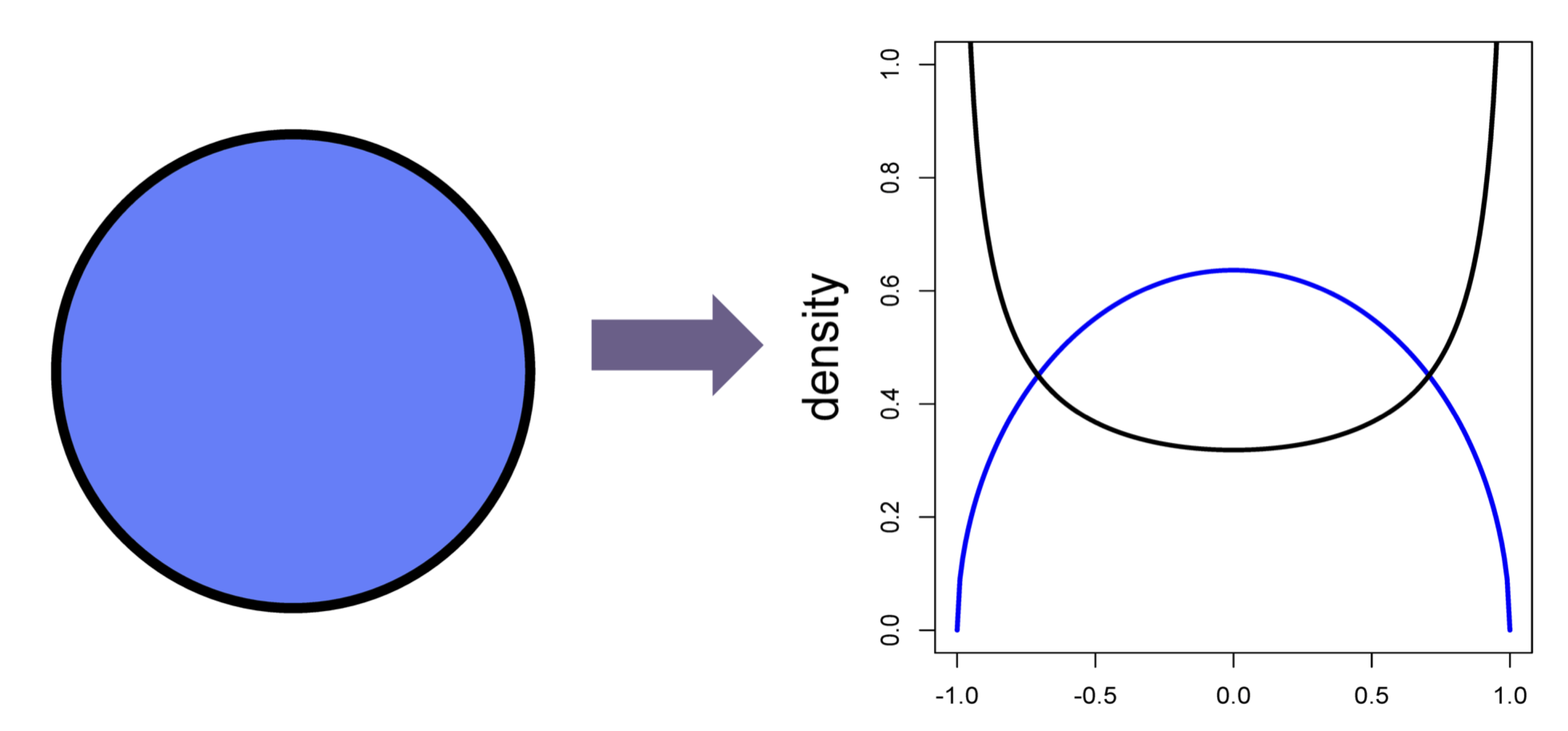}
	\caption{Projections determine distributions! Left: two 2-dimensional distributions, one uniform on the unit circle (black), the other uniform on the unit disk (blue). Right: the corresponding densities after the projecting into a 1-dimensional space.  In fact, any $p$-dimensional distribution is determined by its one-dimensional projections (cf.~Theorem~\ref{thm:Wold}).}
\end{figure}

\renewcommand{\baselinestretch}{1.2}
\normalsize

\clearpage

\section{\sffamily \Large INTRODUCTION} 
\sloppy Modern approaches to data analysis go far beyond what early statisticians, such as Ronald~A.~Fisher, may have dreamt up.  Rapid advances in the way we can collect, store and process data, as well as the value in what we can learn from data, has led to a vast number of innovative and creative new methods.  

Broadly speaking, random projections offer a universal and flexible approach to complex statistical problems. They are a particularly useful tool in large-scale settings, such as high-dimensional classification \citep{DurrantKaban:2013,DurrantKaban:2015,CanningsSamworth:2017}, clustering \citep{Dasgupta:1999,FernBrodley:2013,HeckelTschannenBolskei:2017}, precision matrix estimation \citep{Marzetta:11}, regression \citep{KlankeVijayakumarSchaal:2008,McWilliams:14,Heinzeetal:2016,Thaneietal:2017,Thaneietal:2018,Slawski:2018,DobribanLiu:2019}, sparse principal component analysis \citep{Gataric:2019}, hypothesis testing \citep{Lopes:2011,ShiLuSong:2019}, correlation estimation \citep{Grellmannetal:2016}, dimension reduction \citep{BinghamMannilla:2001,ReeveMuBrown:2018} and matrix decomposition \citep{HalkoMartinssonTropp:2011}.

Random projections are an example of a data perturbation technique. In general, data perturbation refers to an approach in which one does not apply a method directly to the raw data set, but rather looks at a perturbed version (or perhaps many perturbed versions) of the data.  This idea has a long history: perhaps the most well-known example is the Bootstrap. Since Efron coined the term in his seminal 1979 paper \citep{Efron:1979}, the Bootstrap has been extensively studied and developed -- it has received multiple book-length treatments, e.g. \citet{EfronTibshirani:1993}, \citet{ShaoTu:1995} and \citet{DavisonHinkley:1997}; see also the more recent paper \citet{KleinerTalwalkerSurkarJordon:2014} on large-scale applications of the Bootstrap. The Bootstrap works by recalculating many statistics based on different random subsamples (with replacement) of the observations, with the aim of understanding the uncertainty of an estimator.  

In prediction problems, bootstrap aggregation, or \textit{bagging} \citep{Breiman:1996a}, can be used to improve the accuracy of a simple method. By aggregating the results of many predictions based on bootsrapped predictions, one can obtain a final accurate prediction with low variance. See, for instance, \citet{HallSamworth:2005}, \citet{BiauDevroye:2010}, \citet{Biauetal:2010} and \citet{Samworth:2012} who study the properties of the bagged nearest neighbour classifier.  The extremely popular random forests algorithm, neatly combines bagging with classification and regression trees \citep{Breimanetal:1984,Breiman:2001}.

Other, more recent data perturbation techniques include \textit{stability selection} \citep{MeinshausenBuhlmann:2010,ShahSamworth:2013}, which is designed to improve the performance of a variable selection algorithm by aggregating the results of applying a base selection procedure to many subsamples of the data.  \citet{ShahMeinshausen:2014} propose a related method called \textit{random intersection trees}, which aims to find interactions  between the variables in high-dimensional problems. The \textit{knockoff filter} \citep{BarberCandes:2015} guarantees control of the false discovery rate in a variable selection problem, by constructing exchangeable ``knockoff" copies of the features that are independent of the response -- if a variable is not selected before its knockoff copy it is likely to be a false discovery.  Finally, \citet{HintonSKSS:2012} propose a data perturbation method called \textit{dropout}, which aims to prevent a neural network from overfitting -- see also \citet{WagerWangLiang:2013}.

The remainder of this paper focusses on random projection methods in classification problems.  Classification is one of the fundamental problems in statistical learning. In the simple, binary version, we are presented with the task of assigning a test observation to one of two classes, based on a number of training observations from each class.  This problem dates back at least to the aforementioned Fisher, who applied his Linear Discriminant Analysis (LDA) method to identify the species of Iris plants based on measurements of the petal and sepal sizes \citep{Fisher:36}.  Modern applications are seemingly endless; think, for instance, of a spam filter sorting email into the appropriate folders; a driverless car determining whether a hazard is approaching, a doctor classifying tumours in an x-ray; or a smart-watch recognising the wearers activity and adding one (or not) to their step-count for the day.  

Classification has been studied by statisticians, computer scientists, machine learners and AI researchers.  The basic methods include, among hundreds of others, linear discriminant analysis (LDA), its quadratic counterpart (QDA), $k$-nearest neighbours ($k$nn) \citep{FixHodges:1951}, support vector machines (SVM) \citep{Cortes:95}, trees and random forests \citep{Breimanetal:1984,Breiman:2001}, empirical risk minimisation (ERM) \citep{Vapnik:1992}, kernel methods \citep{Marron:1983} and neural networks and deep learners \citep{LeCunBengioHinton:2015}.  Further discussion of these techniques and many others can be found in, for example, \citet{PTPR:96}, \citet{BoucheronBousquetLugosi:2005} and \citet{ESL:09}.  

A common theme in modern applications is the size of the datasets involved -- we often have a huge amount of data.  The term high-dimensional refers to a situation where the number of features $p$ is comparable to or larger than (perhaps much larger than) the total number of observations $n$.  This setting typically leads to problems for existing methods -- the so-called \textit{curse of dimensionality}.  Either we lose statistical accuracy \citep{BickelLevina:2004} or  suffer a prohibitive computational cost. In fact, some methods are simply intractable in high-dimensional settings, for example LDA requires the inverse of a sample covariance matrix, which will be singular if $p > n$; see also \citet[Example~1.2.1]{Wainwright:2019}.  

Recently, there have been a number of proposals aimed at dealing with high-dimensional data in a classification problem -- see, for examples, \citet{Friedman:1989}, \citet{HastieBujaTibshirani:1995},  \citet{Tibshirani:2002}, \citet{Tibshirani:2003},  \citet{FanFan:2008}, \citet{WittenTibshirani:2011} and \citet{FanFengTong:2012}. It is typically assumed in these works that the optimal decision boundary is linear, and only a small proportion of the features are relevant for classification. 

In this article, we see that random projections offer an alternative solution to high-dimensional classification problems. The use of random projections is motivated by two fundamental results.  The first is that a distribution is determined by its one-dimensional projections -- a result sometimes referred to as the Cram\'er--Wold device.
\begin{theorem}
	\label{thm:Wold}
	Suppose $X_1$ and $X_2$ are independent random vectors taking values in $\mathbb{R}^p$. Fix $d \in \{1, \ldots, p\}$.  If $AX_1 =^d A X_2$ for every $A \in \mathbb{R}^{d \times p}$, then $X_1 = ^d X_2$. 	  
\end{theorem} 
The proof of this result is very simple using characteristic functions: For $t \in \mathbb{R}^p$, there exists $s \in \mathbb{R}^d$ and $A \in \mathbb{R}^{d \times p}$ such that $t = A^Ts$. Thus 
	\[
	\mathbb{E}\{\exp(it^TX_1)\} = \mathbb{E}\{\exp(is^T A X_1)\} =  \mathbb{E}\{\exp(is^T A X_2)\} =  \mathbb{E}\{\exp(it^TX_2)\}.
	\]
Heuristically speaking, therefore, we can \emph{learn} a high-dimensional distribution by looking only at its low-dimensional projections.  

Our second motivating result is the Johnson--Lindenstrauss Lemma. This states that a set of arbitrary points in a high-dimensional ambient space can be mapped into a low-dimensional space, while approximately preserving the pairwise distances between the points.  We state the result in a form relating directly to random projections. 
\begin{theorem} 
\label{thm:JL}
	Let $\epsilon, \delta >0$, and for $n \geq 2$, suppose that $x_1, \ldots, x_n$ are distinct vectors in $\mathbb{R}^p$. Fix $d > \frac{16 \log(n/\delta)}{\epsilon^2}$, and  let $\textbf{A}$ be a random projection taking values in $\mathbb{R}^{d \times p}$ with independent $N(0,1/p)$ entries.  Then, with probability greater than $1 - \delta$, we have
	\[
1-\epsilon < \frac{\|\textbf{A}x_j - \textbf{A} x_j\|^2}{\|x_i - x_j\|^2} < 1 + \epsilon, \quad \text{for every $i \neq j \in \{1, \ldots, n\}$.}
\]
\end{theorem} 
The proof of the Johnson--Lindenstrauss Lemma is based on the concentration of $\chi^2$ random variables; more details can be found in, for example, \citet{DasguptaGupta:2002}, \citet{Ailon:2006} and \citet[Example~2.12]{Wainwright:2019}.  The power of this result is perhaps quite striking: notice that the lower bound on the projected dimension $d$ does not depend on the ambient dimension $p$, and scales only logarithmically in the number of data points $n$.  Suppose, for instance, that we have 1000 observations in one million dimensions -- a scale often seen in modern applications -- let $\epsilon = 0.1$ and $\delta = 0.01$, then the lower bound on $d$ is around $18000$. In other words, using just one random projection of the data, we can reduce the dimension by a number of orders of magnitude, while potentially almost preserving the statistical efficiency.  

Many authors have sought to simplify the random projections used in the Johnson--Lindenstrauss Lemma \citep{Achlioptas:2003, LiHastieChurch:2006, LeSarlosSmola:2013}. Moreover, \citet{LarsenNelson:2016} showed that the lower bound on $d$ is optimal. 
    
The remainder of this paper will focus on two approaches to using random projections in classification.   The first we will call \textit{ensemble methods} (cf.~Section~\ref{sec:ensemble}), which typically seek to improve the statistical accuracy of a method, by applying it to many random projections of the data.  The second, which will be referred to as \textit{sketching} (cf.~Section~\ref{sec:sketch}), looks to improve the computational efficiency of an algorithm by first reducing the effective sample size or data dimension using a random projection, with the hope that one does not lose out in terms of statistical accuracy.  We then conclude the paper with a number of discussion points and open problems. First, in the next section, we introduce the statistical framework used throughout the paper.  
   
\section{\sffamily \Large STATISTICAL SETTING}
Let $(X,Y), (X_1, Y_1), \ldots, (X_n, Y_n)$, be independent and identically distributed pairs, taking values in $\mathbb{R}^p \times \{0,1\}$, with distribution $P$.  We observe the \textit{training data} $\mathcal{T}_n = \{(X_1, Y_1), \ldots, (X_n, Y_n) \}$, the \textit{test} point $X$ and would like to predict the class $Y$.  Here $n$ and $p$ will be referred to as the sample size and (ambient) dimension, respectively.  We can characterise the joint distribution $P$ by fixing the marginal $X$ distribution $P_X$ and specifying the regression function $\eta(x) := \mathbb{P}(Y=1 | X= x)$; alternatively, we can fix the marginal $Y$ distribution by specifying prior probabilities $\pi_1 := \mathbb{P}(Y = 1) = 1 - \mathbb{P}(Y = 0) =: 1- \pi_0$, and then generate $X$ according to the class-conditional distribution $X| \{Y = r\} \sim P_r$, for $r = 0,1$.  

A classifier is a (measurable) function $C: \mathbb{R}^{p} \rightarrow \{0,1\}$, with the interpretation that the point $x \in \mathbb{R}^p$ is assigned to the class $C(x)$.  It is useful to let $\mathcal{C}_p$ denote the space of all $p$-dimensional classifiers.  In practice, we construct classifiers based on the training data $\mathcal{T}_n$, which we will typically denote by $C_n: (\mathbb{R}^p \times\{0,1\})^{\otimes n} \rightarrow \mathcal{C}_p$.  In other words, $C_n$ is a rule (or algorithm) that constructs a classifier in $\mathcal{C}_p$ depending on the training data $\mathcal{T}_n$.

We often seek classifiers with low test (or misclassification) error 
\[
R(C) := \int_{\mathbb{R}^p \times \{0,1\}} \mathbbm{1}_{\{C(x) \neq y\}} \, dP(x,y). 
\] 
We write the test error as an integral here to make it clear that we are only averaging over the distribution of the test pair $(X,Y)$. The test error is minimised by the Bayes classifier 
\[
C^{\mathrm{Bayes}}(x) := \biggl\{\begin{array}{l l } 1 & \text{if $\eta(x) \geq 1/2$} \\ 0 & \text{otherwise;}\end{array}
\]
see, for instance, \citet[Theorem 2.1]{PTPR:96}. We have that $R(C^{\mathrm{Bayes}}) = \mathbb{E}[\min \{\eta(X), 1 - \eta(X)\} ].$

In what follows, we will construct classifiers using the (random) training data as well as random projections.  In order to keep track of the different sources of randomness, random projections will be displayed in bold (typically by~$\mathbf{A}, \mathbf{A}_1$, etc.). Fixed, non-random projections will be presented in plain typeface, i.e.~$A$, $A_1$, etc.. We use $\mathbf{E}$ and $\mathbf{P}$ to denote expectation and probability, respectively, taken over the randomness of the projections (conditionally on the training data).   On the other hand, $\mathbb{E}$ and $\mathbb{P}$ are used to refer to expectation and probability, respectively, over all sources of randomness (i.e.~random training data, the random test point, and the random projections).    We will use the convention that $A$ and  $\mathbf{A}$ will take values in $\mathbb{R}^{d \times p}$, and will therefore map a point $x \in \mathbb{R}^p$ to $A x \in \mathbb{R}^d$. Finally, the term \emph{Gaussian random projection} will be used to refer to a projection with independent $N(0, 1/p)$ entries, a \emph{Haar} projection is one uniformly distributed on the set $\{A \in \mathbb{R}^{d \times p}: AA^T = I_{d\times d} \}$, whereas an \emph{axis-aligned} projection has orthonormal rows and one non-zero entry equal to 1 in each row. 

\section{ENSEMBLE METHODS}
\label{sec:ensemble}
Ensemble methods work by aggregating many (typically randomised) estimators.  The intuition is that combining the results of many noisy but unbiased predictions will lead to an unbiased prediction with low variance. The aforementioned bagging procedure is perhaps the most widely used ensemble approach in classification. As noted by Breiman\footnote{\citet[p.~124]{Breiman:1996a}} \emph{``bagging can push a good but unstable procedure towards optimality. On the other hand, it can slightly degrade the performance of stable procedures"}. A similar statement can be made about some methods based on random projections. 

Early empirical work demonstrated the potential power of random projection ensembles: \citet{SchclarRokach:2009} showed that a simple majority vote random projection ensemble classifier was competitive with bagging in some settings.  In the remainder of this section, we provide an overview of some recent random projection based ensemble methods, with a particular aim to summarise the associated theoretical guarantees.   

\subsection{\sffamily \large The random-projection ensemble classifier}
\label{sec:CS17}
The \textit{random-projection ensemble classifier}, proposed in a recent paper by \citet{CanningsSamworth:2017}, works by aggregating the results of applying an arbitrary base classifier to many carefully chosen low-dimensional random projections of the data. It can be seen \textit{``as a general technique for either extending the applicability of an existing method to high dimensions, or simply improving its performance"}.\footnote{\citet[p.~961]{CanningsSamworth:2017}} 

\begin{figure}[h!]
	\centering
	\includegraphics[width=0.9\textwidth]{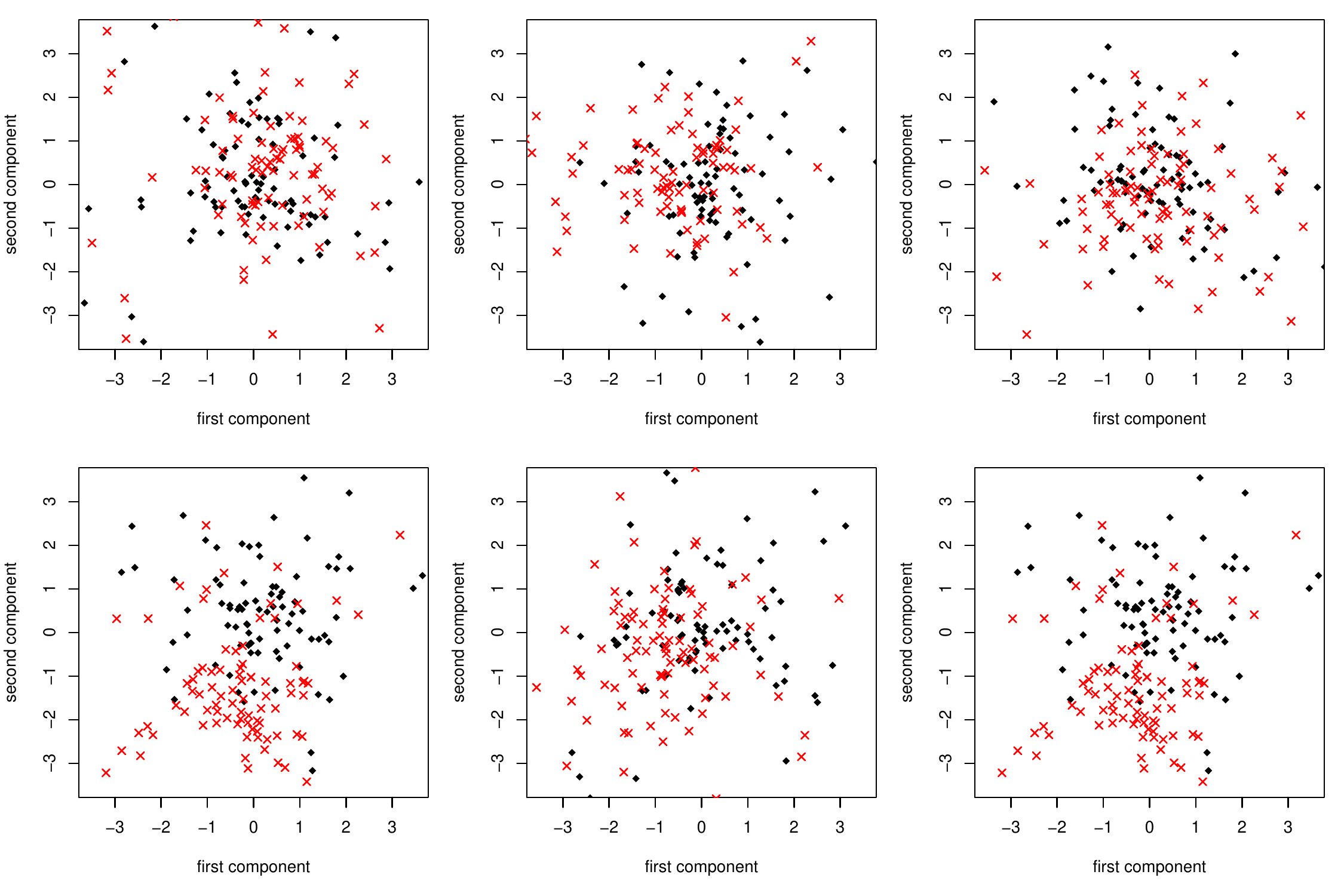}
	\caption{Different two-dimensional projections of $200$ observations in $p = 50$ dimensions.  Top row: three projections drawn from Haar measure; bottom row: the projections with the smallest estimate of test error out of 100 Haar projections for the LDA (left), QDA (middle) and $k$nn (right) base classifiers. Reproduced with permission from \citet[Fig.~1]{CanningsSamworth:2017}.}
	\label{fig:useless}
\end{figure}

One of the key observations is that aggregating the results of applying all the projections is not effective -- indeed, most (low-dimensional) random projections in the high-dimensional setting lead to a random guess -- see Figure~\ref{fig:useless}.   \citet{CanningsSamworth:2017} instead therefore advocate selecting good projections based on an estimate of the test error after applying each one. A second key observation is that combining the results via a simple majority vote is typically not suitable -- indeed the intuition one might have from bagging no longer applies -- the random-projection ensemble classifier instead uses a biased majority vote, with a data-driven voting threshold.   

The random-projection ensemble classifier is given in Algorithm~\ref{algo:RPEnsemble}. We first formally define some notation used in the construction of the classifier.  Let $d \leq p$ (one should think of $d$ being small, $\leq 10$, say), and assume we have a base classifier $C_{n,d} = C_{n,\mathcal{T}_{n,d}}$, which can be constructed from any training sample $\mathcal{T}_{n,d}$ of size $n$ in $\mathbb{R}^d \times \{0,1\}$; thus $C_{n,d}$ is a measurable function from $(\mathbb{R}^d \times \{0,1\})^n \times \mathbb{R}^d$ to  $\{0,1\}$.  Given a projection $A \in \mathbb{R}^{d\times p}$, let $\mathcal{T}_n^A := \{(A X_1,Y_1),\ldots,(AX_n,Y_n)\}$.  The projected data base classifier corresponding to $C_{n,d}$ is $C_n^A: (\mathbb{R}^d \times \{0,1\})^n \times \mathbb{R}^p \rightarrow \{0,1\}$, given by
\[
C_n^A(x) = C_{n,\mathcal{T}_n^A}^A(x) := C_{n,\mathcal{T}_n^A}(Ax).
\]
Note that although $C_n^A$ is a classifier on $\mathbb{R}^p$, the value of $C_n^A(x)$ only depends on $x$ through its $d$-dimensional projection $Ax$.  Now, let ${R}_n^A$ be an estimator of the test error $R({C}_n^A)$ based on $\mathcal{T}_n^A$.  Examples of such estimators include the training error and leave-one-out estimator; see \citet[Section~4]{CanningsSamworth:2017} for more detail. 

\bigskip
\begin{algorithm}[H]
	\caption{\textit{The random-projection ensemble classifier}}
	\SetAlgoLined
	\KwResult{$C_n^{\mathrm{RP}}(x) = \mathbbm{1}_{\{\nu_n^{B_1}(x) \geq  \alpha\}}$}
	\KwData{$\mathcal{T}_n$ and the test point $x\in \mathbb{R}^p$}
	\textbf{Input}: $\alpha \in [0,1]$, $B_1, B_2, d \in \mathbb{N}$, a projected data base classifier $C_{n,d}$\;
	\For{$b_1 = 1, \ldots, B_1$}{ 
		\For{$b_2 = 1, \ldots, B_2$}{
			Generate a Gaussian projection $\textbf{A}_{b_1,b_2}$\;
			Project the training data to give $\mathcal{T}_n^{\mathbf{A}_{b_1,b_2}}$\;
			Estimate $R(C^{\mathbf{A}_{b_1,b_2}}_{n})$ by $R_n^{\mathbf{A}_{b_1,b_2}}$\; 
		}
		Set $\textbf{A}_{b_1} =  \textbf{A}_{b_1,b_2^*(b_1)}$, where $b_2^*(b_1) := \sargmin_{b_2 = 1,\ldots, B_2} \{R_n^{\mathbf{A}_{b_1,b_2}}\}$\;
	}
	Let $\nu_n^{B_1}(x) :=  \frac{1}{B_1}\sum_{b_1=1}^{B_1} \mathbbm{1}_{\{C_n^{\mathbf{A}_{b_1}}(x) = 1 \}}$.
	\label{algo:RPEnsemble} 
\end{algorithm}
\bigskip

Notice the flexibility offered by the framework in Algorithm~\ref{algo:RPEnsemble}. The practitioner may use their favourite base classifier (\citet{CanningsSamworth:2017} study the LDA, QDA and $k$nn base classifiers in detail), any projection distribution and any way of measuring the performance of a projection.  Details of how to choose $\alpha$ depending on the training data are given in \citep[Section~5.2]{CanningsSamworth:2017}.  The method can be implemented using the \texttt{RPEnsemble} \citep{CanningsSamworth2016b} package available from \texttt{CRAN}.

\begin{figure}[h]
	\centering
	\makebox{\includegraphics[width=\textwidth]{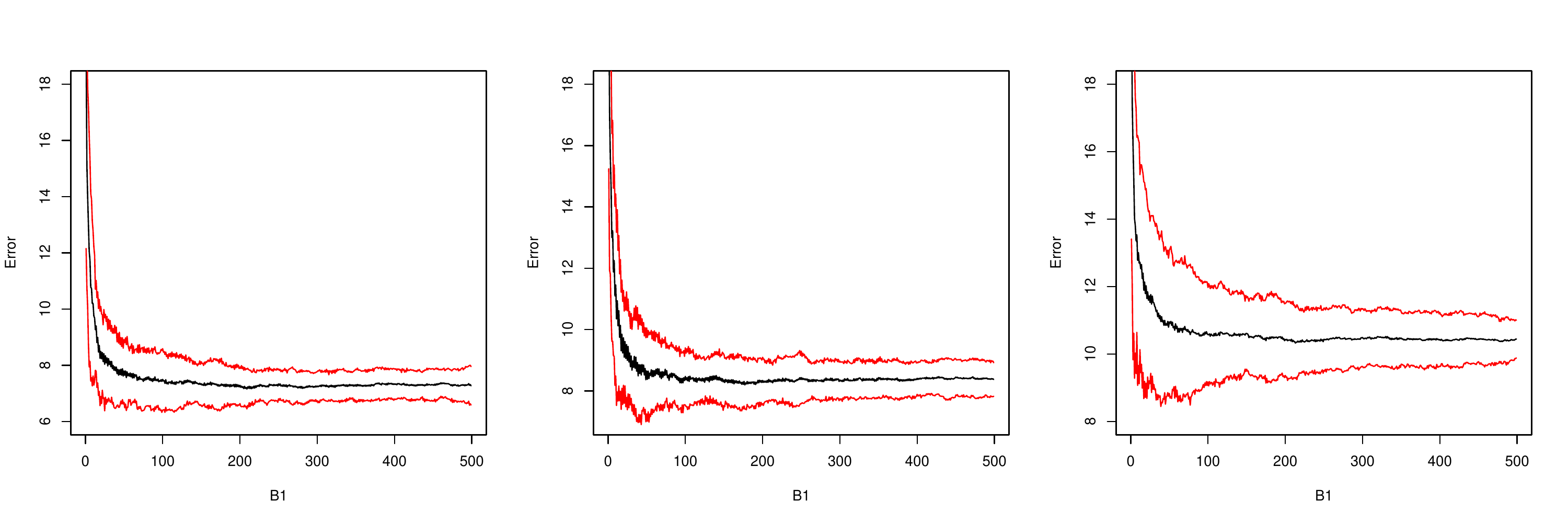}}
	\caption{\label{fig:errors} The average error (black) plus/minus two standard deviations (red) of $C_n^{\mathrm{RP}}$ over 20 sets of $B_1B_2$ projections for $B_1 \in \{2,\ldots, 500\}$ and $B_2 = 50$.  The plots show the test error for one training dataset for the LDA (left), QDA (middle) and $k$nn (right) projected data base classifiers. Reproduced with permission from \citet[Fig.~2]{CanningsSamworth:2017}.}
\end{figure}

The construction in Algorithm~\ref{algo:RPEnsemble} means that the chosen projections $\textbf{A}_{1}, \ldots, \textbf{A}_{B_1}$ depend on the training data, but are in fact conditionally independent given the training data.  This allows for theoretical analysis. Indeed, \citet[Theorem~1]{CanningsSamworth:2017} studies the performance of the algorithm as $B_1$ increases; it is shown that the error of an ensemble using $B_1$ random projections converges to the error of the infinite ensemble at rate $B_1^{-1}$ -- thus one should choose $B_1$ as large as possible up to a computational constraint -- see Figure~\ref{fig:errors} for a numerical demonstration of this.  

The choice of $B_2$ is less straightforward due to potential issue of overfitting \citep[Theorem~3]{CanningsSamworth:2017}. If $B_2$ is too small, then we may only be averaging over noise (cf.~Figure~\ref{fig:useless}). On the other hand if $B_2$ is too large, then we may select a set of projections that are good for the training data, but do not generalise well.  From a practical viewpoint, the default choices of $d = 5$, $B_1 = 500$ and $B_2 = 50$ typically work very well (cf.~Section~\ref{sec:epilepsy}). 

Another main theoretical contribution in \citet{CanningsSamworth:2017} is that, under a low-dimensional structure assumption, the average performance of the random-projection ensemble classifier may be bounded by terms that do not depend on the ambient dimension~$p$ (cf.~\citet[Theorem~3 and Proposition~2]{CanningsSamworth:2017}).  More precisely, the assumption is as follows: for a projection $A$ and $z \in \mathbb{R}^d$, let $\eta^{A}(z) := \mathbb{P}(Y=1 | AX = z)$. Now, suppose that there exists a projection $A^*  \in \mathbb{R}^{d \times p}$, for which 
\begin{equation}
\label{eq:A*}
P_X\bigl(\{x \in \mathbb{R}^p : \eta(x) \geq 1/2\} \Delta \{x \in \mathbb{R}^p : \eta^{A^*}(A^*x) \geq 1/2\} \bigr) = 0,
\end{equation}
where $B\Delta C = (B \cap C^c) \cup (B^c \cap C)$ denotes the symmetric difference of two sets $B$ and $C$  (cf.~\citet[Assumption~3]{CanningsSamworth:2017}). This condition can be seen as a generalisation of those typically used in the high-dimensional classification literature: if the Bayes decision boundary is linear, then~\eqref{eq:A*} holds with $d =1$; under the common sparse signal assumption that only $d$ of the features are relevant for classification, \eqref{eq:A*} holds with an \textit{axis-aligned} choice of $A^*$;  finally, \eqref{eq:A*} holds under the sufficient dimension reduction assumption that $Y$ is conditionally independent of $X$ given $A^*X$ \citep{Cook:1998}.

The computational cost of the random-projection ensemble classifier is discussed in \citet[Section~5.1]{CanningsSamworth:2017}.  One of the appealing features of the method is its compatibility with parallel computing -- we can simultaneously compute the projected data base classifier for each of the $B_1B_2$ projections.      

There was a stimulating and constructive discussion of the paper.  A number of methodological variations were proposed. These included how to generate the random projections, to subsequently assess each projection's performance, and then how to aggregate the results.  The original paper focussed on Haar and Gaussian distributed projections.  Alternatives include using \textit{axis-aligned} projections, which are well-suited to the ultrahigh-dimensional setting, or \textit{very sparse} random projections \citep{LiHastieChurch:2006} -- see also \citet{MylavarapuKaban:2013} for a direct comparison of random projections versus random feature selection.  Another suggestion was to sequentially update the projections, attempting to improve the predictions each time. While intuitively appealing, \citet{CanningsSamworth:2017} found that, in fact, (as observed by Breiman) having a diverse set of projections is desirable -- some discussants even suggested to enforce some orthogonality constraint. Typically, however, in high dimensions two projections chosen as described in Algorithm~\ref{algo:RPEnsemble} are close to orthogonal anyway. 

Other discussants suggested alternative ways to aggregate the results.  Many proposed a weighted combination similar to that used in boosting \citep{FreundSchapire:1996,FreundSchapire:1999}; or to use a blending/stacking method \citep{Wolpert:1992,Breiman:1996b}, which involves applying a classifier that uses the predictions from the base classifiers themselves as new features. 

\citet{BlaserFryzlewicz:2015} investigate a related method using random rotations as opposed to projections. In their work a classifier suited to high-dimensional data is applied after each rotation.   In a follow up paper, \citet{BlaserFryzlewicz:2019} advocate selecting good rotations based on their complexity, where simpler learners are preferred.   \citet{Gul:2016} propose an ensemble method based on applying the $k$-nearest neighbour classifier to subsamples of the training data -- they randomly choose subsets of both the features and the observations. This process is repeated many times and the top performing projections (measured on an out-of-bag sample) are retained.  The results of applying the $k$nn classifier with the chosen samples are then combined to construct the final classifier.  \citet{Khan:2015} propose a method based on tree classifiers. \citet{XiaoWang:2017} study an ensemble of randomly chosen linear base classifiers, and provide a bound on the performance of their method based on the VC-dimension. They show empirically that it is competitive with random kitchen sinks \citep{RahimiRecht:2007} and Adaboost \citep{FreundSchapire:1999}.

There is also some recent general theoretical work on understanding ensemble methods.  \citet{Lopes:2019b} derives the rate at which the test error of a finite ensemble approaches its infinite simulation counterpart. \citet{Lopes:2019a} proposes a bootstrap method to approximate the variance of an ensemble, with a view to ascertain how large an ensemble is needed. 

In summary, the random-projection ensemble classifier offers a general approach to high-dimensional statistical problems.  At a high level, just three key ingredients are required: (i) a suitable low-dimensional method for the problem at hand; (ii) a measure of the relative performance after applying each projection; and (iii) an effective aggregation procedure.   \citet{Gataric:2019} introduce a new method for sparse principal component analysis based on this framework -- in their work, the target is to obtain a low-dimensional projection of the data that explains the greatest proportion of the population variance.  Since the components in the projection are assumed to be sparse, it is preferable to use axis-aligned projections, as opposed to Gaussian projections.  Very recently, \citet{Anderlucci:2019} applied this framework in the unsupervised clustering problem, where a Gaussian mixture model assumption is used in order to asses the quality of each projection, and a technique known as consensus clustering is used to aggregate the results. 

\subsection{\sffamily \large Model based ensembles}
\label{sec:modelensemble}
Other works have exploited the use random projections to directly estimate the model parameters in a high-dimensional classification problem.  This is the setting investigated in \citet{DurrantKaban:2015} (see also \citet{Marzetta:11}), where multiple random projections of the data are used to estimate the high-dimensional precision matrix in LDA.

Suppose, for simplicity, that $X | \{Y = r\} \sim N_p(\mu_r, \Sigma)$, for $r = 0,1$, where $\mu_0, \mu_1 \in \mathbb{R}^p$, and $\Sigma$ is a $p \times p$ covariance matrix common to both classes.  The Bayes classifier in this case is
\begin{equation}
\label{eq:LDABayes}
C^{\mathrm{LDA-Bayes}}(x) = \biggl\{ \begin{array}{ll} 1  & \mbox{\quad if $\log\bigl(\frac{\pi_1}{\pi_0}\bigr) +  \bigl(x - \frac{{\mu}_1 + {\mu}_0}{2}\bigr)^T {\Sigma}^{-1} ({\mu}_1 - {\mu}_0) \geq 0$}
\\ 0 & \mbox{\quad otherwise.} \end{array}
\end{equation}
Its risk can be expressed in terms of $\pi_0, \pi_1$, and the squared Mahalanobis distance $\Delta^2 = (\mu_1 - \mu_0)^T \Sigma^{-1} (\mu_1 - \mu_0)$ between the classes 
\[
R(C^{\mathrm{LDA-Bayes}}) = \pi_0\Phi\Bigl(\frac{1}{\Delta} \log\Bigl(\frac{\pi_1}{\pi_0}\Bigr) - \frac{\Delta}{2}\Bigr) +  \pi_1\Phi\Bigl(\frac{1}{\Delta}\log\Bigl(\frac{\pi_0}{\pi_1}\Bigr) - \frac{\Delta}{2}\Bigr),
\]
where $\Phi$ denotes the standard normal distribution function. 

The LDA classifier is constructed by substituting training data estimates of $\pi_0, \pi_1, \mu_0, \mu_1,$ and $\Sigma$ in to \eqref{eq:LDABayes}.  These are given by $\hat{\pi}_r = \frac{1}{n} \sum_{i = 1}^{n} \mathbbm{1}_{\{Y_i = r\}}$, $\hat{\mu}_r = \frac{1}{n_r} \sum_{i = 1}^{n} X_i \mathbbm{1}_{\{Y_i = r\}}$, and 
\[
\hat{\Sigma} = \frac{1}{n-2}\sum_{i=1}^n\sum_{r =0}^1 (X_i - \hat{\mu}_r)(X_i - \hat{\mu}_r)^T \mathbbm{1}_{\{Y_i = r\}}.
\]
As mentioned in the introduction, if $p > n$, then $\hat{\Sigma}$ will be singular, and LDA is intractable is its vanilla form. 

\citet{DurrantKaban:2015} advocate estimating $\Sigma$ using random projections.  For $B \in \mathbb{N}$, let $\mathbf{A}_1, \ldots, \mathbf{A}_B$ be independent Gaussian random projections taking values in $\mathbb{R}^{d \times p}$.  Then let
\[
\label{eq:LDAEns}
C_n^{\mathrm{LDA-Ens}}(x) = \biggl\{ \begin{array}{ll} 1  & \mbox{\quad if $\log\bigl(\frac{\hat{\pi}_1}{\hat{\pi}_0}\bigr) + \frac{1}{B}\sum_{b = 1}^{B} \bigl(x - \frac{{\hat{\mu}}_1 + {\hat{\mu}}_0}{2}\bigr)^T \mathbf{A}_b^T\bigl(\mathbf{A}_b \hat{\Sigma} \mathbf{A}_b^T\bigr)^{-1} \mathbf{A}_b (\hat{\mu}_1 - \hat{\mu}_0) \geq 0$}
\\ 0 & \mbox{\quad otherwise.} \end{array}
\]
In other words, the ensemble uses $\tilde{\Sigma}_B^{-1} = \frac{1}{B}\sum_{b = 1}^{B}  \mathbf{A}_b^T\bigl(\mathbf{A}_b \hat{\Sigma} \mathbf{A}_b^T\bigr)^{-1} \mathbf{A}_b$ as an estimate of $\Sigma^{-1}$. Note that the terms in the sum $\mathbf{A}_b \hat{\Sigma} \mathbf{A}_b^T$ are invertible almost surely as long as $d < \min\{n,p\}$.   

Now, by the law of large numbers, if $\mathbf{E}\{\mathbf{A}_1^T\bigl(\mathbf{A}_1 \hat{\Sigma} \mathbf{A}_1^T\bigr)^{-1} \mathbf{A}_1 \}$ exists, then we have that
\[
\bigl(x - \frac{{\hat{\mu}}_1 + {\hat{\mu}}_0}{2}\bigr)^T \tilde{\Sigma}_B^{-1}  (\hat{\mu}_1 - \hat{\mu}_0)  \stackrel{\mathrm{a.s.}}{\rightarrow}  \bigl(x - \frac{{\hat{\mu}}_1 + {\hat{\mu}}_0}{2}\bigr)^T \mathbf{E}\Bigl\{\mathbf{A}_1^T\bigl(\mathbf{A}_1 \hat{\Sigma} \mathbf{A}_1^T\bigr)^{-1} \mathbf{A}_1 \Bigr\} (\hat{\mu}_1 - \hat{\mu}_0).
\]
\citet[Theorem 3.2]{DurrantKaban:2015} derive a bound on the test error of an LDA classifier that uses the converged ensemble precision matrix $\mathbf{E}( \tilde{\Sigma}_B^{-1})$. The bound depends on how well $\mathbf{E}( \tilde{\Sigma}_B^{-1})$ approximates $\Sigma^{-1}$, as well as the squared Mahalonobis distance $\Delta$ and the balance between the two class sizes.  The accuracy of the precision matrix estimate itself $\tilde{\Sigma}_B^{-1}$ (with finite $B$) was further investigated in \citet{Kaban:2017}.

\subsection{\sffamily \large The epileptic seizure recognition data set}
\label{sec:epilepsy}
We now demonstrate the utility of the methods described in this section with a brief numerical study.   The epileptic seizure recognition data set \citep{Andrzejaketal:2001} available from the UCI Machine Learning repository\footnote{\url{https://archive.ics.uci.edu/ml/datasets/Epileptic+Seizure+Recognition}} contains 11500 observations of a 179-dimensional EEG recording.  Associated with each observation is a label in $\{1,\ldots, 5\}$ corresponding to whether the patient was experiencing an epileptic seizure or not.  We simplify the problem by combining the four ``no seizure" classes, so that the task is to predict $Y \in \{0,1\}$, where class 0 and class 1 correspond to ``no seizure'' and ``seizure", respectively.    In the resulting dataset, there are 9200 observations in class 0 and 2300 in class 1.  

\begin{figure}[h]
	\centering
	\makebox{\includegraphics[width=0.5\textwidth]{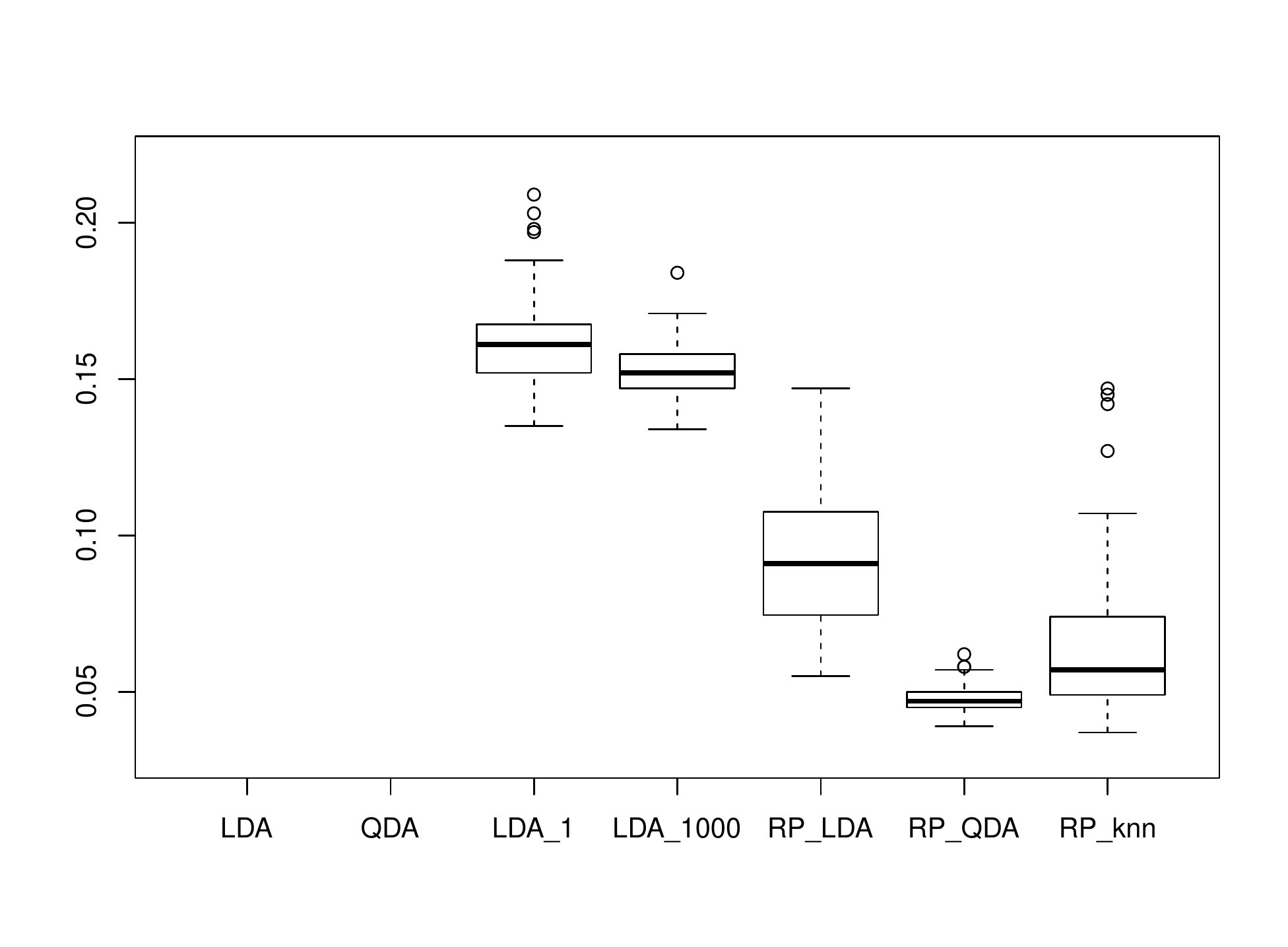}\includegraphics[width=0.5\textwidth]{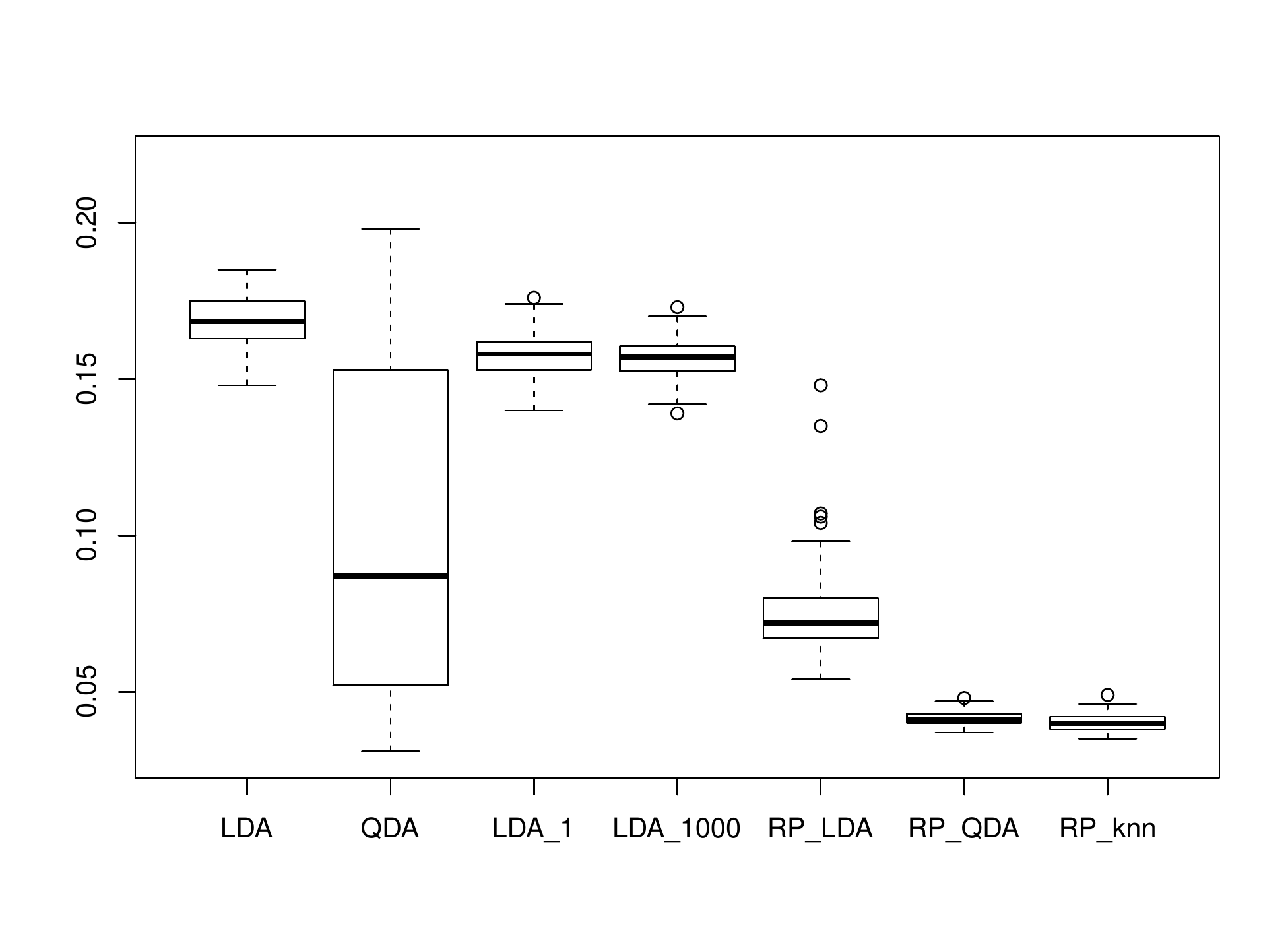}}
	\caption{\label{fig:epilepsy} The estimated test errors for the experiments described in Section \ref{sec:epilepsy} using the epileptic seizure recognition data set. Left panel: $n = 100$. Right panel: $n = 1000$.}
\end{figure}

To assess the performance of the classifiers, we take a random sample of size 1000 to use as a test set (this remains fixed throughout the study) and our experiments are repeated 100 times on different (randomly chosen) training samples.  There are two studies: one with $n = 100$ and one with $n = 1000$.   We compare seven classifiers: (vanilla) LDA, (vanilla) QDA, two based on $C_n^{\mathrm{LDA-Ens}}$, and the random-projection ensemble classifier with three different choices of base classifiers.  For $C_n^{\mathrm{LDA-Ens}}$, we set $d = \frac{1}{2} \min\{n-2,p\}$ as recommended in \citet{DurrantKaban:2015}, and we use an ensemble of $B = 1000$ Gaussian random projections (LDA\_1000) -- for comparison we also include the results when just one projection is used (LDA\_1).  For the random-projection ensemble classifiers, we use the LDA, QDA, and $k$nn base methods and the default parameters recommended in \citet{CanningsSamworth:2017}, that is $d= 5$, $B_1 = 500$ and $B_2 = 50$.  The voting cutoff $\alpha$ was chosen using the method described in \citet[Section 5.2]{CanningsSamworth:2017}.  These methods are denoted by RP\_LDA, RP\_QDA and RP\_knn in Figure~\ref{fig:epilepsy}.

In Figure~\ref{fig:epilepsy} we present boxplots of the test errors for the 100 repetitions of each experiment. First note that, for $n =100$, the LDA and QDA classifiers are intractable since $n < p$.  In fact, for $n = 1000$, there are 5 out of the 100 experiments where QDA is still intractable, since there were fewer than 179 observations in the minority class in those cases.   We see that the LDA ensemble method of \citet{DurrantKaban:2015} offers a tractable version of the LDA classifier, but it is not particularly effective in this problem.  The random-projection ensemble classifier with the QDA base classifier is very accurate for both sample sizes and the $k$nn base classifier gives the best results when $n = 1000$. 

\section{SKETCHING AND HASHING}
\label{sec:sketch}
The aim of sketching and hashing is to reduce the complexity of a problem, by reducing the (effective) sample size or dimension, respectively, while approximately preserving the statistical efficiency.  These techniques can often lead to a huge computational saving -- in fact, in some cases, we may not have sufficient disk space to store the raw data, and therefore some form of sketching may be required.   In contrast to the previous section on ensemble methods, typically only one projection or sketch is used to train the classifier.  

Perhaps the simplest random sketching approach is to subsample the observations -- suppose, for example, that we have a huge number of observations ($10^6$, say), but are interested in a straightforward problem, such as LDA.  If the data dimension is low, then we will perhaps obtain sufficiently accurate results with around 1000 observations; including the full dataset in the estimation of the parameters in LDA will only give minor improvements.   As a result we do not need to store the full dataset, and our estimation procedure will be much faster.  With this approach the data dimension is unchanged and it is unlikely to be successful if $p$ is large.  Of course, with a large amount of data available, it is also likely that a more sophisticated approach than LDA will be possible. 

Other works have investigated sketching techniques that involve premultiplying the $n \times p$ data matrix $(X_1, \ldots, X_n)^T$ and the $n\times 1$ vector of responses (or classes) $(Y_1, \ldots, Y_n)^T$ by a random $\mathbf{\Omega} \in \mathbb{R}^{m \times n}$ projection.  Again, like subsampling,  the dimension of the problem stays the same, but the effective sample size may be reduced significantly.  This technique has received a fair amount of attention in the context of kernel ridge regression \citep{YangPilanciWainwright:2017,AvronClarksonWoodruff:2017,DobribanLiu:2019}, but comparatively little in classification problems.  

There are a number of works that advocate applying an existing classifier after projecting the features into a lower dimensional space.  Typically, the idea in these problems is to reduce the dimension, and thus the computational cost, while preserving performance guarantees using an argument similar to the Johnson--Linderstrauss Lemma (cf.~Theorem~\ref{thm:JL}).  Note that, in contrast to Section~\ref{sec:CS17}, where low-dimensional projections were used (i.e.~$d \leq 10$), for the Johnson--Lindenstrauss Lemma to be effective, the projection dimension should grow with the logarithm of the sample size $n$. For instance, it is often shown that under some condition on the dimension of the image space of the map that, with high-probability (over the randomness in the projection), the error of the classifier trained on the projected data  is close to that which could be obtained by training the classifier in the ambient high-dimensional space.  

This approach has been studied in combination with Fisher's linear discriminant analysis \citep{DurrantKaban:2010,DurrantKaban:2012,ElkhalilKammounCalderbankAlNaffouriAlouini:2019,SkubalskaRafajlowicz:2019}. Recall the class-conditional Gaussian setting introduced in Section~\ref{sec:modelensemble}. Let $\mathbf{A}$ be a Gaussian random projection and define
  \[
  \label{eq:LDARP}
  C_n^{\mathrm{LDA-}\mathbf{A}}(x) = \biggl\{ \begin{array}{ll} 1  & \mbox{\quad if $\log\bigl(\frac{\hat{\pi}_1}{\hat{\pi}_0}\bigr) + \bigl(x - \frac{{\hat{\mu}}_1 + {\hat{\mu}}_0}{2}\bigr)^T \mathbf{A}\bigl(\mathbf{A}\hat{\Sigma} \mathbf{A}^T\bigr)^{-1} \mathbf{A}(\hat{\mu}_1 - \hat{\mu}_0) \geq 0$}
  \\ 0 & \mbox{\quad otherwise.} \end{array}
  \]
\citet[Theorem~4.8]{DurrantKaban:2012} provide a bound on the average (over the projection) test error of $C_n^{\mathrm{LDA-}\mathbf{A}}$.  A similar result was shown in \citet{DurrantKaban:2013} for a classifier based on linear empirical risk minimisation.  One of the key aspects of these works is the so-called \emph{flipping probability} \citep[Theorem 3.2]{DurrantKaban:2013}, which specifies the chance that the label assigned to a point in the ambient $p$-dimensional is ``flipped" (from zero to one or vice-versa) after applying a random projection.

Other works in this direction focus on alternative base methods, for instance, the $k$-nearest neighbour classifier \citep{Ailon:2006,Kaban:2015,ReeveBrown:2017} and support vector machines \citep{RahimiRecht:2007,Paul:2012}.  \citet{XieLiZhangWang:2016} investigate combining random projection techniques with other dimension reduction methods, such as principal component analysis. 

In some settings it is in fact possible to exactly encode a high-dimensional dataset via a low-dimensional representation.  Indeed, suppose that the feature vectors are high-dimensional, binary and sparse -- i.e. $X$ takes values in $\{0,1\}^p$, but only a small proportion of the features are non-zero for each observation.  \citet{ShahMeinshausen:2018} propose an approach to large scale classification and regression in this context, based on $b$-bit min-wise hashing \citep{LiKonig:2011}.  They show how the min-wise hashing technique can be combined with logistic regression in order to give improved computational and statistical efficiency.  

Finally, we mention that some hashing and sketching techniques are designed to guarantee privacy -- by applying a non-invertable map (or projection) to the data, we can ensure that any sensitive information is hidden -- for some examples in this direction, see \citet{KenthapadiKorolovaMirinovMishra:2013} and \citet{Upadhyay:2013}.

\section{DISCUSSION} 
Despite the large body of work mentioned in this review, the use of random projections in classification problems (and indeed in wider statistical problems) is perhaps still in its early stages.  A number of practical considerations remain.  Perhaps at the forefront of those are general concerns about randomised methods -- for instance, two different practitioners may obtain different results using the same method, simply by using different initial randomisation seeds.  That being said, ensemble approaches partially derandomise procedures, and the huge popularity of methods like random forests suggests that practitioners are often happy to overlook this issue. 

Many random projection based approaches are so-called \emph{black box} methods -- they may classify accurately, but offer limited interpretation as to how a decision was made.  In some applications this is not an issue. Think, for example, of an email spam filter, where, if an email is designated to the spam folder, we're not interested in why that decision was made.  On the other hand, suppose a doctor is using a randomised algorithm to help diagnose a disease, it is of limited practical use if the classifier simply produces a yes or no answer (unless it is perfectly accurate).  

One way to aid interpretability is to provide a relative ranking of the importance of each of the features in the model. There is some numerical work in this direction, for instance, \citet[Section~10]{Breiman:2001} proposes a variable importance measure for the random forest algorithm.  Moreover, for the random-projection ensemble classifier, the chosen projections provide a natural way of ranking the features.  There is, however, relatively little understanding of the precise theoretical properties of such approaches. 

Further considerations include testing the robustness of such methods -- what if the data is noisy or missing?  There has been a fair amount of work recently on label noise  \citep{FrenayKaban:2014,FrenayVerleysen:2014}.  Simple methods such as $k$-nearest neighbours and support vector machines have been shown to be robust to label noise \citep{CanningsFanSamworth:2019}.  It is less clear, however, how more sophisticated methods, such as those based on random projections, will be affected by noise.   

There are many other open questions remaining on the use of random projections in statistics. First, there are computational and statistical trade-offs that are not precisely understood. How about optimality -- what can be learnt (in a minimax sense) from random projections of the data?  Finally, while a distribution is determined by the distributions of its projections (cf.~Theorem~\ref{thm:Wold}), and we perhaps have a good understanding of how well we can approximate the low-dimensional distributions from projected data, it is not understood how this translates to learning the properties of the high-dimensional distribution.


\begin{thebibliography}{99}
\bibitem[Achlioptas(2003)]{Achlioptas:2003} Achlioptas, D. (2003) Database-friendly random projections: Johnson--Lindenstrauss with binary coins. \newblock \textit{J. Comp. and Sys. Sci.}, \textbf{4}, 671--687. 

\bibitem[{Ailon and Chazelle(2006)}]{Ailon:2006} Ailon, N. and Chazelle, B. (2006) Approximate nearest neighbours and the fast Johnson--Lindenstrauss transform. \newblock \emph{Proceedings of the Symposium on Theory of Computing}, 557--563. 

\bibitem[Anderlucci et al.(2019)]{Anderlucci:2019} Anderlucci, L., Fortunato, F. and Montanari, A. (2019) High-dimensional clustering via random projections. \newblock \textit{Preprint}, \texttt{ArXiv:1909.10832}. 

\bibitem[Andrzejak et al.(2001)]{Andrzejaketal:2001} Andrzejak, R.~G., Lehnertz, K., Rieke, C., Mormann, F., David, P., and Elger, C.~E. (2001) Indications of nonlinear deterministic and finite dimensional structures in time series of brain electrical activity: Dependence on recording region and brain state.  \newblock \textit{Phys. Rev. E}, \textbf{64}, 061907. 


\bibitem[Avron et al.(2017)]{AvronClarksonWoodruff:2017} Avron, H., Clarkson, K.~L. and Woodruff, D. P. (2017)  Faster kernel ridge regression using sketching and preconditioning. \newblock \textit{SIAM J. Matrix Anal. Appl.}, \textbf{38}, 1116--1138. 

\bibitem[Barber and Cand\`{e}s(2015)]{BarberCandes:2015} Barber, R.~F. and, Cand\`{e}s, E.  (2015) Controlling the false discovery rate via knockoffs. \newblock \emph{Ann. Statist.}, \textbf{43}, 2055--2085. 

\bibitem[Biau et al.(2010)]{Biauetal:2010} Biau, G., C\'{e}rou, F. and Guyader, A. (2010) On the rate of convergence of the bagged nearest neighbor estimate. \newblock \emph{J. Mach. Learn. Res.}, \textbf{11}, 687--712. 

\bibitem[Biau and Devroye(2010)]{BiauDevroye:2010} Biau, G. and Devroye, L. (2010) On the layered nearest neighbour estimate, the bagged nearest neighbour estimate and the random forest method in regression and classification. \newblock \textit{J. Multivariate Analysis}, \textbf{101}, 2499--2518. 

\bibitem[Bickel and Levina(2004)]{BickelLevina:2004} Bickel, P. and Levina, E. (2004) Some theory for Fisher's linear discriminant function, `naive Bayes', and some alternatives when there are more variables than observations. \newblock \textit{Bernoulli}, \textbf{10}, 989--1010. 

\bibitem[Bingham and Mannilla(2001)]{BinghamMannilla:2001} Bingham, E. and Mannilla, H. (2001) Random projection in dimensionality reduction: applications to image and text data. \newblock \textit{Proc. of the Seventh ACM SIGKDD International Conference on Knowledge Discovery and Data Mining}, 245--250.   

\bibitem[{Blaser and Fryzlewicz(2015)}]{BlaserFryzlewicz:2015} Blaser, R. and Fryzlewicz, P. (2015) Random rotation ensembles. \newblock \emph{J. Mach. Learn. Res.}, \textbf{17}, 1--26.

\bibitem[{Blaser and Fryzlewicz(2019)}]{BlaserFryzlewicz:2019}Blaser, R. and Fryzlewicz, P. (2019) Regularizing axis-aligned ensembles via data rotations that favor simpler learners. \newblock \url{http://stats.lse.ac.uk/fryzlewicz/rre/regsim.pdf}.

\bibitem[Boucheron et al.(2005)]{BoucheronBousquetLugosi:2005} Boucheron, S., Boousquet, O. and Lugosi, G. (2005) Theory of classification: a survey of some recent advances. \textit{ESAIM: PS}, \textbf{9}, 323--375. 

\bibitem[{Breiman(1996{\natexlab{a}})}]{Breiman:1996a} Breiman, L. (1996{\natexlab{a}}) Bagging Predictors. \newblock \emph{Machine Learning}, \textbf{24}, 123--140. 

\bibitem[{Breiman(1996{\natexlab{b}})}]{Breiman:1996b} Breiman, L. (1996{\natexlab{b}}) Stacked regressions. \newblock \emph{Machine Learning}, \textbf{24}, 49--64. 

\bibitem[{Breiman(2001)}]{Breiman:2001} Breiman, L. (2001)  Random Forests. \newblock \emph{Machine Learning}, \textbf{45}, 5--32. 

\bibitem[{Breiman et~al.(1984)Breiman, Friedman, Stone and Olshen}]{Breimanetal:1984} Breiman, L., Friedman, J., Stone, C.~J. and Olshen, R.~A. (1984) \emph{Classification and Regression Trees}. \newblock {Chapman and Hall}, New York.

\bibitem[Cannings et al.(2019)]{CanningsFanSamworth:2019} Cannings, T. I., Fan, Y. and Samworth, R.~J. (2019) Classification with imperfect training labels. \newblock \textit{Biometrika}, to appear.

\bibitem[{Cannings and Samworth(2016)}]{CanningsSamworth2016b} Cannings, T. I. and Samworth, R. J. (2016) \texttt{RPEnsemble}: Random projection ensemble classification. \newblock \texttt{R} package, v. 0.3. \url{https://cran.r-project.org/web/packages/RPEnsemble/index.html}

\bibitem[{Cannings and Samworth(2017)}]{CanningsSamworth:2017} Cannings, T. I. and Samworth, R. J. (2017) Random-projection ensemble classification.  \newblock \emph{J. Roy. Statist. Soc., Ser. B. (with discussion)}, \textbf{79}, 959-1035. 

\bibitem[Cook(1998)]{Cook:1998} Cook, R.~D. (1998) \textit{Regression Graphics: Ideas for Studying Regressions through Graphics}. \newblock Wiley, New York. 

\bibitem[{Cortes and Vapnik(1995)}]{Cortes:95} Cortes, C. and Vapnik, V. (1995) Support-vector networks. \newblock \emph{Machine Learning}, \textbf{20}, 273--297. 

\bibitem[Davison and Hinkley(1997)]{DavisonHinkley:1997} Davison, A.~C.  and Hinkley, D.~V. (1997) \textit{Bootstrap Methods and their Applications.} Cambridge University Press, Cambridge, United Kingdom.

\bibitem[{Dasgupta(1999)}]{Dasgupta:1999} Dasgupta, S. (1999) Learning mixtures of Gaussians. \newblock \emph{Proc. 40th Annual Symposium on Foundations of Computer Science}, 634--644.

\bibitem[{Dasgupta and Gupta(2002)}]{DasguptaGupta:2002} Dasgupta, S. and Gupta, A. (2002) An elementary proof of the Johnson--Lindenstrauss Lemma. \newblock \emph{Random Struct. Alg.},  \textbf{22}, 60--65. 

\bibitem[{Devroye et~al.(1996)Devroye, Gy\"orfi and Lugosi}]{PTPR:96} Devroye, L., Gy\"orfi, L. and Lugosi, G. (1996) \emph{A Probabilistic Theory of Pattern Recognition}. \newblock {Springer}, New York.

\bibitem[Dobriban and Liu(2019)]{DobribanLiu:2019} Dobriban, E. and Liu, S. (2019) Asymptotics for sketching in least squares regression. \newblock \textit{NeurIPS2019}, to appear.

\bibitem[{Durrant and Kab\'{a}n(2010)}]{DurrantKaban:2010} Durrant, R.~J. and Kab\'{a}n, A. (2010) Compressed Fisher Linear Discriminant analysis: classification of randomly projected data. \newblock In \emph{Proc. 16th ACM SIGKDD Conference, KDD2010}. 

\bibitem[{Durrant and Kab\'{a}n(2012)}]{DurrantKaban:2012} Durrant, R.~J. and Kab\'{a}n, A. (2012) A tight bound on the performance of Fishers linear discriminant analysis in randomly projected data spaces. \newblock \emph{Pattern Recongition Letters}, \textbf{33}, 911--919. 

\bibitem[{Durrant and Kab\'{a}n(2013)}]{DurrantKaban:2013} Durrant, R.~J. and Kab\'{a}n, A. (2013) Sharp generalization error bounds for randomly-projected classifiers.
\newblock \emph{J. Mach. Learn. Res. W\&CP}, \textbf{28}, 693--701. 

\bibitem[{Durrant and Kab\'{a}n(2015)}]{DurrantKaban:2015} Durrant, R.~J. and Kab\'{a}n, A. (2015) Random projections as regularizers: learning a linear discriminant from fewer observations than dimensions. \newblock \emph{Machine Learning}, \textbf{99}, 257--286. 

\bibitem[Efron(1979)]{Efron:1979} Efron, B. (1979) Bootstrap methods: another look at the Jackknife. \textit{Ann. Statist.}, \textbf{7}, 1--26. 

\bibitem[Efron and Tibshirani(1993)]{EfronTibshirani:1993} Efron, B. and Tibshirani, R.~J. (1993) \textit{An Introduction to the Bootstrap}. Chapman and Hall, CRC, Boca Raton.

\bibitem[Elkhalil et al.(2019)]{ElkhalilKammounCalderbankAlNaffouriAlouini:2019}  Elkhalil, K., Kammoun, A., Calderbank, R.,  Al-Naffouri, T.~Y. and Alouini, M.-S. (2019) Asymptotic performance of linear discriminant analysis with random projections. \newblock In \textit{IEEE International Conference on Acoustics, Speech and Signal Processing (ICASSP 2019)}. 

\bibitem[{Fan and Fan(2008)}]{FanFan:2008} Fan, J. and Fan, Y. (2008) High-dimensional classification using features annealed independence rules. \newblock \emph{Ann. Statist.}, \textbf{36}, 2605--2637. 

\bibitem[{Fan et~al.(2012)Fan, Feng and Tong}]{FanFengTong:2012}Fan, J., Feng, Y. and Tong, X. (2012) A road to classification in high dimensional space: the regularized optimal affine discriminant. \newblock \emph{J. Roy. Statist. Soc., Ser. B.}, \textbf{72}, 745--771. 

\bibitem[Fern and Brodley(2013)]{FernBrodley:2013} Fern, X.~Z. and Brodley, C.~E. (2003) Random projection for high dimensional data clustering: A cluster ensemble approach. \newblock  \emph{ICML-2003}.

\bibitem[{Fisher(1936)}]{Fisher:36} Fisher, R.~A. (1936) The use of multiple measurements in taxonomic problems. \newblock \emph{Annals of Eugenics}, \textbf{7}, 179--188. 

\bibitem[{Fix and Hodges(1951)}]{FixHodges:1951} Fix, E. and Hodges, J.~L. (1951) Discriminatory analysis -- nonparametric discrimination: Consistency properties.
\newblock Technical Report 4, USAF School of Aviation Medicine, Randolph Field, Texas.

\bibitem[Freund and Schapire(1996)]{FreundSchapire:1996} Freund, Y. and Schapire, R. E. (1996) A decision theoretic generalisation of on-line learning and an application to boosting. \newblock \textit{Journal of Computer and System Sciences}, \textbf{55}, 119--139. 
	
\bibitem[Freund and Schapire(1999)]{FreundSchapire:1999} Freund, Y. and Schapire, R.~E. (1996) A short introduction to boosting. \newblock \textit{Journal of Japanese Society for Artificial Intelligence}, \textbf{14}, 771--780.
	
\bibitem[{Fr\'{e}nay \& Kab\'{a}n(2014)}]{FrenayKaban:2014} Fr\'{e}nay, B. \& Kab\'{a}n, A. (2014) A comprehensive introduction to label noise. \newblock \textit{Proc. Euro. Sym. Artificial Neural Networks}, 667--676.

\bibitem[{Fr\'{e}nay \& Verleysen(2014)}]{FrenayVerleysen:2014} Fr\'{e}nay, B. \& Verleysen, M. (2014) Classification in the presence of label noise: a survey. \newblock \textit{IEEE Trans. on NN and Learn. Sys.}, \textbf{25},  845--869. 
		
\bibitem[{Friedman(1989)}]{Friedman:1989} Friedman, J. (1989) Regularized discriminant analysis. \newblock \emph{J. Amer. Statist. Assoc.}, \textbf{84}, 165--175. 

\bibitem[{Gataric et al.(2019)}]{Gataric:2019} Gataric, M. , Wang, T. and Samworth, R.~J. (2019) Sparse principal component analysis via axis-aligned random projections. \newblock \emph{arXiv e-prints}, \texttt{1712.05630}. 

\bibitem[Grellman et al.(2016)]{Grellmannetal:2016} Grellmann, C., Neumann, J., Bitzer, S., Kovacs, P., T\"{o}njes, A.,Westlye, L.~T., Andreasson, O.~A., Stumvoll, M., Villringer, A. and Horstmann, A. (2016) Random projection for fast and efficient multivariate correlation analysis of high-dimensional data: a new approach. \newblock \textit{Frontiers in Genetics}, \textbf{7}, 102. 

\bibitem[{Gul et~al.(2016)Gul, Perperoglou, Khan, Mahmoud, Miftahuddin, Adler and Lausen}]{Gul:2016} Gul, A., Perperoglou, A., Khan, Z., Mahmoud, O., Miftahuddin, M., Adler, W. and Lausen, B. (2016) Ensemble of a subset of $k$NN classifiers. \newblock \emph{Adv. Data Anal. Classif.}, 1-14. 

\bibitem[Halko et al.(2011)]{HalkoMartinssonTropp:2011} Halko, N., Martinsson, P.~G. and Tropp, J.~A. (2011) Finding structure with randomness: Probabilistic algorithms for constructing approximate matrix decompositions. \newblock \textit{SIAM Rev.}, \textbf{53}, 217-288.  

\bibitem[{Hall and Samworth(2005)}]{HallSamworth:2005}Hall, P. and Samworth, R.~J. (2005) Properties of bagged nearest neighbour classifiers. \newblock \emph{J. Roy. Statist. Soc., Ser. B.}, \textbf{67}, 363-379. 

\bibitem[Hastie et al.(1995)]{HastieBujaTibshirani:1995} Hastie, T., Buja, A. and, Tibshirani, R. (1995) Penalized discriminant analysis. \newblock \textit{Ann. Statist.}, \textbf{23}, 73--102. 

\bibitem[{Hastie et~al.(2009)Hastie, Tibshirani and Friedman}]{ESL:09}Hastie, T., Tibshirani, R., and Friedman, J. (2009) \emph{The Elements of Statistical Learning: Data Mining, Inference, and Prediction.} \newblock Springer Series in Statistics (2nd ed.). \newblock Springer, New York.

\bibitem[Heckel et al.(2017)]{HeckelTschannenBolskei:2017} Heckel, R., Tschannen, M. and B\"{o}lcskei, H. (2017) Dimensionality-reduced subspace clustering. \newblock \textit{Information and Inference}, \textbf{6}, 246-283. 

\bibitem[{Heinze et~al.(2016)}]{Heinzeetal:2016} Heinze, C., McWiliams, B.,  Meinshausen, N. (2016) DUAL-LOCO: distributing statistical estimation with random projections. \newblock \emph{AISTATS2016}.

\bibitem[Hinton et al.(2012)]{HintonSKSS:2012} Hinton, G.~E., Srivastava, N., Krizhevsky, A., Sutskever, I. and Salakhutdinov, R.~R. (2012) Improving neural networks by preventing co-adaptation of feature detectors. \newblock \textit{Preprint}, \texttt{Arxiv:1207.0580}.

\bibitem[{Kab\'{a}n(2015)}]{Kaban:2015} Kab\'{a}n, A. (2015) A new look at nearest neighbours: identifying benign input geometries via random projections. \newblock \textit{ACML15}, 65--80.

\bibitem[{Kab\'{a}n(2017)}]{Kaban:2017} Kab\'{a}n, A. (2017) On compressive ensemble induced regularisation: how close is the finite ensemble precision matrix to the infinite ensemble? \newblock \textit{Proc. 28th Intern. Conf. Algo. Learn. Th., PMLR}, \textbf{76}, 617-628.

\bibitem[Kenthapadi et al.(2013)]{KenthapadiKorolovaMirinovMishra:2013}  Kenthapadi, K.,  Korolova, A., Mironov, I. and Mishra, N. (2013) Privacy via the Johnson--Lindenstrauss transform. \newblock \textit{Journal of Privacy and Confidentiality}, \textbf{5},  39--71. 

\bibitem[{Khan et~al.(2015)Khan, Gul, Mahmoud, Miftahuddin, Perperoglou, Adler and Lausen}]{Khan:2015} Khan, Z., Gul, A., Mahmoud, O., Miftahuddin, M., Perperoglou, A., Adler, W. and Lausen, B. (2015) An ensemble of optimal trees for class membership probability estimation. \newblock In: Wilhelm, A., Kestler, H. A. (eds.), \emph{Analysis of Large and Complex Data, European Conference on Data Analysis, Bremen, July, 2014. Series: Studies in Classification, Data Analysis, and Knowledge Organization}. Springer-Verlag, Berlin.

\bibitem[Klanke et al.(2008)]{KlankeVijayakumarSchaal:2008} Klanke, S., Vijayakumar, S. and Schaal, S. (2008) A library for locally weighted projection regression. \newblock \textit{J. Mach. Learn. Res.}, \textbf{9}, 623--626.

\bibitem[Kleiner et al.(2014)]{KleinerTalwalkerSurkarJordon:2014} Kleiner, A., Talwalkar, A., Surkar, P. and Jordan, M. I. (2014) A scalable bootstrap for massive data. \textit{J. Roy. Statist. Soc., Ser. B}, \textbf{76}, 795--816.

\bibitem[{Larsen and Nelson(2016)}]{LarsenNelson:2016}Larsen, K. G. and Nelson, J. (2016) The Johnson--Lindenstrauss lemma is optimal for linear dimensionality reduction.
\newblock \emph{43rd International Colloquium on Automata, Languages and Programming}, \textbf{82}, 1-11.

\bibitem[{Le et~al.(2013)}]{LeSarlosSmola:2013}Le, Q., Sarlos, T. and Smola, A. (2013) Fastfood --- approximating kernel expansions in loglinear time.
\newblock \emph{J. Mach. Learn. Res.}, W \& CP \textbf{28}, 244--252. 

\bibitem[LeCun et al.(2015)]{LeCunBengioHinton:2015} LeCun, Y., Bengio, Y. and Hinton, G. E. (2015) Deep learning. \newblock \textit{Nature}, \textbf{521}, 436--444.

\bibitem[Li et~al.(2006)]{LiHastieChurch:2006} Li, P.,  Hastie, T. and Church, K. (2006) Very sparse random projections. \newblock In \textit{Proceedings of the 12th ACM SIGKDD international conference on Knowledge discovery and data mining},  287--296.

\bibitem[Li and K\"{o}nig(2011)]{LiKonig:2011}  Li, P. and K\"{o}nig, A. (2011) Theory and applications of b-bit minwise hashing. \newblock \textit{Communications of the ACM}, \textbf{54}, 101--109.

\bibitem[{Lopes(2019{\natexlab{a}})}]{Lopes:2019a}Lopes, M. (2019{\natexlab{a}}) Estimating the algorithmic variance of randomized ensembles via the bootstrap.  \newblock \emph{Ann. Statist}, to appear.

\bibitem[{Lopes(2019{\natexlab{b}})}]{Lopes:2019b}Lopes, M. (2019{\natexlab{b}}) Estimating a sharp convergence bound for randomized ensembles.  \newblock \emph{J. Statist. Plan. Inf.}, to appear.

\bibitem[{Lopes et al.(2011)}]{Lopes:2011}Lopes, M., Jacob, L. and Wainwright, M.~J. (2011) A more powerful two-sample test in high dimensions using random projection.
\newblock \emph{Advances in Neural Information Processing Systems (NIPS)}.

\bibitem[Marron(1983)]{Marron:1983} Marron, J.~S. (1983) Optimal rates of convergence to Bayes risk in nonarametric discrimination. \newblock \textit{Ann. Statist.}, \textbf{11}, 1142--1155. 
 
\bibitem[{Marzetta et~al.(2011)Marzetta, Tucci and Simon}]{Marzetta:11}Marzetta, T., Tucci, G. and Simon, S. (2011) A random matrix-theoretic approach to handling singular covariance estimates. \newblock \emph{IEEE Trans. Inf. Th.}, \textbf{57}, 6256--6271.

\bibitem[{McWilliams et~al.(2014)}]{McWilliams:14} McWilliams, B., Heinze, C., Meinshausen, N., Krummenacher, G. and Vanchinathan, H.~P. (2014) LOCO: distributing ridge regression with random projections. \newblock \emph{NIPS workshop on Distributed Machine Learning and Matrix Computations}.

\bibitem[{Meinshausen and B\"uhlmann(2010)}]{MeinshausenBuhlmann:2010}Meinshausen, N. and B\"uhlmann, P. (2010) Stability selection. \newblock \emph{J. Roy. Statist. Soc., Ser. B (with discussion)}, \textbf{72}, 417--473.

\bibitem[Mylavarapu and Kab\'{a}n(2013)]{MylavarapuKaban:2013} Mylavarapu, S. and Kab\'{a}n, A. (2013) Random projections versus random feature selection for classification of high dimensional data. \newblock \textit{Proceedings of the UK Workshop on Computational Intelligence (UKCI 2013)}, 305--312.


\bibitem[Paul et al.(2012)]{Paul:2012} Paul, S., Boutsidis, C., Magdon-Ismail, M. and Drineas, P. (2012) Random projections for support vector machines. \newblock \textit{Proc. 16th Internat. Conf. on Artificial Intelligence and Statistics, AISTATS2013}, 498--506.

\bibitem[Rahimi and Recht(2007)]{RahimiRecht:2007} Rahimi, A. and Recht, B. (2007) Random features for large-scale kernel machines. \newblock \textit{Adv. in Neural Inf. Proc. Sys. NIPS}, \textbf{20}, 1--8. 

\bibitem[Reeve and Brown(2017)]{ReeveBrown:2017} Reeve, H.~R. and Brown, G. (2017) Minimax rates for cost-sensitive learning on manifolds with approximate nearest neighbours. \textit{Proc. Machine Learning Research, Algorithmic Learning Theory}, \textbf{1}, 1--45.
 
 \bibitem[Reeve et al.(2017)]{ReeveMuBrown:2018} Reeve, H.~R., Mu, T. and Brown, G. (2018) Modular dimensionality reduction. \textit{European Conference on Machine Learning, ECML/PKDD}, 605--619. 
 
\bibitem[Samworth(2012)]{Samworth:2012} Samworth, R.~J. (2012) Optimal weighted nearest neighbour classifiers. \newblock \emph{Ann. Statist.}, \textbf{40}, 2733--2763.

\bibitem[Schclar and Rokach(2009)]{SchclarRokach:2009} Schclar, A. and Rokach, L. (2009) Random projection ensemble classifiers. \newblock \textit{ICEIS 2009: Enterprise information systems}, 309--316. 
 
\bibitem[{Shah and Meinshausen(2014)}]{ShahMeinshausen:2014} Shah, R.~D. and Meinshuasen, N. (2014) Random intersection trees. \newblock \emph{J. Mach. Learn. Res.}, \textbf{15}, 629--654. 

\bibitem[{Shah and Meinshausen(2018)}]{ShahMeinshausen:2018}Shah, R.~D. and Meinshuasen, N. (2018) On $b$-bit min-wise hashing for large-scale regression and classification with sparse data. \newblock \emph{J. Mach. Learn. Res.}, \textbf{18}, 1--42. 

\bibitem[{Shah and Samworth(2013)}]{ShahSamworth:2013}Shah, R. D. and Samworth, R. J. (2013) Variable selection with error control: another look at stability selection.
\newblock \emph{J. Roy. Statist. Soc., Ser. B}, \textbf{75}, 55--80. 

\bibitem[Shao and Tu(1995)]{ShaoTu:1995} Shao, J. and Tu, D. (1995) \textit{The Jackknife and Bootstrap}. \newblock Springer, New York.

\bibitem[Shi et al.(2019)]{ShiLuSong:2019} Shi, C., Lu, W. and Song, R. (2019) A sparse random projection-based test for overall qualitative treatment effects. \newblock \textit{J. Amer. Statist. Assoc.}, to appear.

\bibitem[Skubalska-Rafaj\l{}owicz(2019)]{SkubalskaRafajlowicz:2019} Skubalska-Rafaj\l{}owicz, E. (2019) Stability of random-projection based classifiers. The Bayes error perspective. \newblock \textit{SMSA 2019: Stochastic Models, Statistics and their Applications}, 121-130.

\bibitem[Slawski(2018)]{Slawski:2018} Slawski, M. (2018) On principal components regression, random projections, and column sampling. \newblock \textit{Elec. J. Statist.}, \textbf{12}, 3673--3712.

\bibitem[Thanei et al.(2017)]{Thaneietal:2017} Thanei, G.-A., Heinze, C. and Meinshausen, N. (2017) Random projections for large-scale regression. \newblock \emph{Big Data and Complex Analysis}, 51--68. 

\bibitem[Thanei et al.(2018)]{Thaneietal:2018} Thanei, G.-A., Meinshausen, N. and Shah, R.~D. (2018) The xyx algorithm for fast interaction search in high-dimensional data. \newblock \emph{J. Mach. Learn. Res.}, \textbf{19}, 1--42. 

\bibitem[{Tibshirani et~al.(2002)Tibshirani, Hastie, Narisimhan and Chu}]{Tibshirani:2002}Tibshirani, R., Hastie, T., Narisimhan, B. and Chu, G. (2002) Diagnosis of multiple cancer types by shrunken centroids of gene expression. \newblock \emph{Proceedings of the Natural Academy of Science, USA},  \textbf{99}, 6567--6572.

\bibitem[{Tibshirani et~al.(2003)Tibshirani, Hastie, Narisimhan and Chu}]{Tibshirani:2003} Tibshirani, R., Hastie, T., Narisimhan, B. and Chu, G. (2003) Class prediction by nearest shrunken centroids, with applications to {DNA} microarrays. \newblock \emph{Statist. Science}, \textbf{18}, 104--117.

\bibitem[Upadhyay(2013)]{Upadhyay:2013}  Upadhyay, J. (2013) Random projections, graph sparsification, and differential privacy. \newblock \textit{International Conference on the Theory and Application of Cryptology and Information Security, ASIACRYPT 2013}, 276--295.

\bibitem[Vapnik(1992)]{Vapnik:1992} Vapnik, V. (1992) Principles of risk minimization for learning theory. \newblock \textit{Adv. Neu. Inf. Proc. Sys}, 1992 

\bibitem[Wager et al.(2013)]{WagerWangLiang:2013} Wager, S., Wang, S. and Liang, P.~S. (2013) Dropout training as adaptive regularization. \newblock \textit{Advances in Neural Information Processing Systems}, 351--359.

\bibitem[{Wainwright(2019)}]{Wainwright:2019} Wainwright, M.~J. (2019) \textit{High-dimensional statistics: a non-asymptotic viewpoint.} Cambridge University Press, Cambridge, United Kingdom.

\bibitem[{Witten and Tibshirani(2011)}]{WittenTibshirani:2011} Witten, D.~M. and Tibshirani, R. (2011) Penalized classification using Fisher's linear discriminant.\newblock \emph{J. Roy. Statist. Soc., Ser. B.}, \textbf{73}, 753--772.

\bibitem[Wolpert(1992)]{Wolpert:1992} Wolpert, D. (1992) Stacked generalisations. \newblock \textit{Neural Networks}, \textbf{5}, 241--259.

\bibitem[Xiao and Wang(2017)]{XiaoWang:2017} Xiao, Q. and Wang, Z. (2017) Ensemble classification based on random linear base classifiers. \newblock In \textit{IEEE International Conference on Acoustics, Speech and Signal Processing (ICASSP 2017)}. 

\bibitem[Xie et al.(2016)]{XieLiZhangWang:2016} Xie, H., Li, J., Zhang, Q. and Wang, Y. (2016) Comparison among dimensionality reduction techniques based on Random Projection for cancer classification. \newblock \textit{Computational Biology and Chemistry}, \textbf{65}, 165--172.

\bibitem[Yang et al.(2017)]{YangPilanciWainwright:2017} Yang, Y., Pilanci, M.  and Wainwright, M.~J. (2017) Randomized sketches for kernels: Fast and optimal non-parametric regression. \newblock \textit{Ann. Statist.}, \textbf{45}, 991--1023.

\section*{\sffamily \Large FURTHER READING} 
\subsection*{\sffamily \large General topics}
\bibitem[B\"{u}hlmann and van de Geer(2011)]{BvdG:2011} B\"{u}hlmann, P. and van de Geer, S. (2011) \textit{Statistics for high-dimensional data.} \newblock Springer, 2011.

\bibitem[{Kab\'{a}n(2019)}]{Kaban:2019} Kab\'{a}n, A. (2019) Dimension-free error bounds from random projections. \newblock \textit{AAAI19}.

\bibitem[{Vershynin(2012)}]{Vershynin2012}Vershynin, R. (2012) Introduction to the non-asymptotic analysis of random matrices. \newblock In \emph{Compressed Sensing} (Eds. Y. C. Eldar and G. Kutyniok), pp. 210--268.  Cambridge University Press, Cambridge.

\subsection*{\sffamily \large The knockoff filter} 
\bibitem[Barber and Cand\'{e}s(2016)]{BarberCandes:2016} Barber, R.~F. and, Cand\`{e}s, E.  (2016) A knockoff filter for high-dimensional selective inference. \newblock \emph{Ann. Statist.}, \textbf{47}, 2504--2537. 

\bibitem[Barber et al.(2019)]{Barber:2019} Barber, R.~F., Cand\`{e}s, E.  and Samworth, R.~J.(2019) Robust inference with knockoffs. \newblock \emph{Ann. Statist.}, to appear. 

\bibitem[Cand\`{e}s et al.(2018)]{Candesetal:2018} Cand\`{e}s, E.~J., Fan, Y., Janson, L., Lv, J. (2018) Panning for gold: `model-X' knockoffs for high-dimensional controlled feature selection. \newblock \emph{J. Roy. Statist. Soc., Ser. B}, \textbf{80}, 551-577.

\subsection*{\sffamily \large Large-scale optimisation}
\bibitem[{Sanyang and Kaban(2019)}]{SangyangKaban:2019} Sanyang, M.~L. and Kaban, A. (2019) Large scale estimation of distribution algorithm with adaptive heavy tailed random projection ensembles. \newblock \textit{Journal of Computer Science and Technology}, to appear.

\subsection*{\sffamily \large Random kernel expansions}
\bibitem[Bach(2017)]{Bach:2017} Bach, F. (2017) On the equivalence between kernel quadrature rules and random feature expansions. \textit{J. Mach. Learn. Res.}, \textbf{18}, 1--38. 

\bibitem[Rahimi and Recht(2008)]{RR:2008} Rahimi, A. and Recht, B. (2008) Weighted sums of random kitchen sinks: replacing minimization with randomization in learning. \newblock In \textit{Advances in Neural Information Processing Systems}.

\end{thebibliography}
\end{document}